\documentclass{article}
\usepackage{amssymb}
\usepackage{amsmath} 

\begin{document}
\numberwithin{equation}{section}

\title{Asymptotic Dynamics of Ripples 
\footnote{ PACS \#  03.40.Kf, 47.35.+i \hfill E-mail: manna@lpm.univ-montp2.fr }}
\author{ M. A. Manna  \\
{\em Physique Math\'ematique et Th\'eorique, CNRS-UMR5825,}\\
 Universit\'e Montpellier 2, 34095 MONTPELLIER (France)}
\maketitle

\begin{abstract} A new nonlinear equation governing asymptotic dynamics of
ripples is derived   by using a short wave perturbative expansion on a
generalized version of the Green-Naghdi system. It admits peakon
solutions with amplitude, velocity and width in interrelation and static
compacton solutions with amplitude and width in interrelation. Short
wave pattern formation is shown to result from a balance between linear
dispersion and  nonlinearity.
\end{abstract}

\section{Introduction}
\label{sec:intro}

Since the classical works of Boussinesq \cite{Bouss} and Korteweg and de
Vries\cite {kadeve}, nonlinear evolution of long waves of small amplitude in
shallow fluids has been widely studied. The asymptotic dynamics (in space and
time) is by now well understood and represented by a large number of model
equations, among which the different versions of the Boussinesq equations
\cite {Whitham}, the Korteweg-de Vries (KdV), the
Benjamin-Bona-Mahoney-Peregrini (BBMP) \cite{Bona}, 
 and the
Camassa-Holm \cite{camassa} equations.

On the contrary, {\it short waves} have been studied very little and only a few
results are known on their asymptotic dynamics. The main purpose of 
this paper
is to study nonlinear short surface waves in fluids, the {\em ripples}, shown
to build up as a result of superposition of two surface motions: an oscillatory
flow and a laminar flow. The oscillatory flow corresponds to mechanical
perturbations which propagate like a wave.  The laminar flow may be created in
many ways: by the action of an external wind, or in a two-layer liquid where
the upper one displaces  particles belonging to the upper surface of
the lower fluid, or by an external electric field if the surface particles are
charged, etc...  

Here we study an ideal fluid (inviscid, incompressible and
without surface tension) in which surface displacement can be achieved through
the action of a steady external wind directed parallel to the water surface.
A surface wind on a lake which produces surface flow is an ideal
physical environment conducive to the above mentioned phenomenon.

\section{Model equation for ripples propagation}

Let's consider $u(x,t)$ which represents at
time $t$ an unidirectional surface wave propagation in the
$x$ direction of a fluid medium which is involved in a flow on a
large scale. This large scale flow is a superposed motion of the
surface which moves under the action of the wind with a velocity
$c_{0}$ in relation to the bulk. 
Using perturbative methods we will show that ripples can
propagate and obey the nonlinear equation
\begin{equation}\label{mannaeq}u_{xt} = -\frac{3g}{hc_{0}}u - uu_{xx} +
(u_x)^2
\end{equation}
where subscripts denote partial derivatives, $h$ is the unperturbed
initial depth and $g$ is the acceleration of gravity. 
The equation (\ref{mannaeq})  has several types of interesting
solutions. The peakon solution is (using the formula 
$(1 - \partial_{zz} )e^{-\vert z \vert} = 2\delta(z)$, \cite{camassa}) 
\begin{equation}\label{picon}
u = -\alpha\lambda^{2}\exp(-|\frac{x + \alpha\lambda^{2}t}{
\lambda}|), \,\,\,\,\,\,\alpha =-\frac{3g}{hc_{0}},   
\end{equation} 
where the width $\lambda$ is a free parameter. Unlike the peakon of  Camassa-Holm
equation, the amplitude $(-\alpha\lambda^{2})$, the velocity
$(\alpha\lambda^{2})$ and the width $(\lambda)$ are interrelated.  

The static compacton solution  is 
\begin{equation}\label{compacton}  
 u = -8 \alpha \lambda^{2}\cos^2(
\frac{x}{4\lambda}),\,\,\,\, |\frac{x}{\lambda}| \leq 2\pi , 
\end{equation}
and $ u = 0$ otherwise. Unlike the compacton solution recently introduced
and investigated by Roseneau and Hyman \cite {roseneau},  this solution presents
a dependence between width and amplitude.  
Solutions (\ref{picon}) is a coherent struture analogous
to the solitary wave of KdV.  Moreover, contrary to KdV, (\ref{mannaeq})
possess a plane monochromatic wave solution of arbitrary amplitude $A$
\begin{equation}
u =  A\exp i(kx-\Omega t)
\label{ondeplane}
\end{equation}
with the dispersion relation $\Omega (k) = 3g/khc_0$ identical to the one
of the linearized system  \cite{observation}. Moreover this dispersion relation
is a function of $k$ which has a good behavior in the short wave limit.

Despite the fact that solution (\ref{compacton}) is unnatural in the 
physics under consideration, it
shows that (\ref {mannaeq}) is an adequate mathematical tool to
modelize statics patterns in nature.

The underlying mechanism responsible for structural stability of solutions of
the Camassa-Holm equation or the KdV equation with nonlinear dispersion is the
balance between nonlinear dispersion, nonlinear convection and nonlinearity. 
Indeed these equations are nonlinear evolution equations without linear
dispersion (the plane wave is not a solution of the linear associated evolution
equations). However the exibited solutions of (\ref{mannaeq}) come 
from the
balance between linear dispersion and two nonlinear terms (in the case
of (\ref{picon})) and only nonlinearity (in the case of
(\ref{compacton})). An essential point, which remains to be proved  by
 numerical or analytical methods is: {\it under what conditions do the 
solution (\ref{picon}) dominates the
initial value problem of equation (\ref{mannaeq})?}.  Another important open
problem is the {\it structural stability} of (\ref{picon}) and (\ref{compacton})
. Here we only show that the plane wave (\ref{ondeplane}) is
unstable.  

\section{Physical context: modified Green-Naghdi system of
equations.}


We consider an inviscid, incompressible, homogeneous fluid with density
$\sigma$. Let the particles of this continuum medium be identified by a fixed
rectangular Cartesian system of center $O$ and axes $(x_1, x_2, x_3) = (x, z,
y)$ with  $Oz$ the upward vertical direction. We assume symmetry in $y$ and we
will only consider a sheet of fluid in the $xz$ plane. This fluid sheet is
moving in a domain with a rigid botton at $z = 0$ and an upper free surface at
$z = \phi(x,t)$. The vector velocity is $ \vec v = (v_1, v_2) = (u,w)$ for $0
\leq z < \phi(x,t)$. Thanks to homogenity and incompressibility the continuity
equation reduces to
\begin{equation}\label{continuity}
\vec \nabla \cdot \vec v = u_x(x, z, t) + w_z(x, z, t) = 0,
\,\,\, 0 \leq z \leq \phi(x,t),
\end{equation}
where $\vec \nabla = (\partial_x,\partial_z)$. The Euler
equations of motions (law of conservation of momentun) of a fluid
under gravity $g$ and for $0 \leq z \leq \phi(x,t)$ are 
\begin{equation}\label{euler1}
\sigma \dot u (x, z, t) = -p^*_x(x, z, t), 
\end{equation}
\begin{equation}\label{euler2}
\sigma \dot w (x, z, t) = -p^*_z(x, z, t) - g\sigma,
\end{equation} 
where $p^*(x, z, t)$ is the pressure and a superposed dot denotes the material
derivative for $x, z$ fixed. 

We complete now the fundamental equations of
continuity and momentun conservation with appropiate kinematic and 
dynamics
boundaries conditions. Let $S(x,z,t)$ be the interface between the inviscid
fluid sheet and the air (external medium). We represented  $S(x,z,t)$ by the
(classical) equation
\begin{equation}\label{surfaceS}
S(x,z,t) = \phi(x,t) - z = 0.
\end{equation}
The kinematic condition is that the normal velocity of the surface
$S(x,z,t)$ must be equal to the velocity of the fluid sheet normal
to the surface. The normal velocity of the surface is \cite{lamb}  
\begin{equation}\label{norvel}		
-\frac { S_t }{ \parallel\vec\nabla S \parallel } , 
\end{equation}
while the velocity on the surface $z = \phi(x,t)$ is
\begin{equation}\label{velalasurf}
\vec v = ( u + c_0, w), \quad z = \phi(x,t)\ ,
\end{equation}
whose normal component is
\begin{equation}\label{velnoraS}
\vec v \cdot \frac {\vec\nabla S}{\parallel\vec\nabla S \parallel}\ .  
\end{equation}
From (\ref{norvel}), (\ref{velalasurf}) and (\ref{velnoraS}) the
kinematic conditions read 
\begin{equation}\label{kincond}
\phi_t + u\phi_x - w + c_0\phi_x = 0, \quad z = \phi . 
\end{equation}

The equation (\ref{velalasurf}) leading to (\ref{kincond}) lies at the heart of
our approach. It points out that an external agent - wind in our case - drives
the particles belonging to $S(x, z, t)$. The motion is uniform, of value $c_0$
and in the $x$ direction only. So $S(x, z, t)$ is a surface of discontinuity
for the $u$ component of $\vec v$, which experiences finite jumps of value
$c_0$ in $z = \phi$. This wind induced motion of the fluid's surface is found
in particular in the hydrodynamics of lakes \cite {lakes}. From a theoretical
point of view this phenomenon was studied in the ray tracing theory by
Lighthill in \cite {Lighthill} where an equation analogous to (\ref{kincond})
was derived.  

At the surface of the sheet $z = \phi$, there is a constant normal pressure
$p_0 $. At the bed $z = 0$, there is an unknown pressure $p^*(x, 0, t)$ and the
normal fluid velocity is zero: $w = 0$. Several types of approximations can be
used in order to solve this wave problem. Here we adopt an approximation in the
velocity field introduced by Green, Laws and Naghdi in \cite{Naghdi1} and based
on the monumental work of Naghdi \cite{Naghdi2}.  We assume that $u$ is
independent of $z$. This is equivalent to considering the vertival component
$w$ as a linear function of $z$. This simple and realistic assumption enables us
 to
satisfy exactly the equation of incompressibility and the boundary condition at
the bed. Hence $u= u(x, t)$ and from (\ref{continuity}) we have
\begin{equation}\label{eqpourpsi} 
w = z\xi(x, t)\ ,\quad \xi(x, t) = -u_x(x, t).
\end{equation}
Now we integrate (\ref{euler1}) in the variable $z$ from $z = 0$ to $z = \phi$
(the integration is granted by the Riemann's condition of integrability). 
The result is
\begin{equation}\label{euler1integ}
\sigma(u_t + uu_x)\phi = -p_x,
\end{equation}
where we use $ \dot u = u_t + uu_x$ and 
\begin{equation}\label{presion} 
p(x, t) = \int_0^{\phi(x, t)}p^*(x, z, t)dz - p_0 \phi(x, t). 
\end{equation}
Next, we multiply equation (\ref{euler2}) by $z$ and integrate from
 $z = 0$ to $z = \phi$ which yields
\begin{equation}\label{euler2integ}
\sigma( \xi^2 + \dot\xi)\frac{\phi^3}{3} + \sigma g\frac{\phi^2}{2} =p.
\end{equation}
The pressure $p$ can be eliminated using (\ref{euler2integ}) in  
(\ref{euler1integ}) and, with the help of (\ref{eqpourpsi}) we
eventually obtain 
\begin{eqnarray}\label{GN1}
u_t + uu_x + g\phi_x =&& \phi \phi_x (u_{xt} + uu_{xx} - (u_x)^2) + 
\nonumber\\
&&\frac{\phi^3}{3}(u_{xt} + uu_{xx} - (u_x)^2)_x 
\end{eqnarray}
The remaining upper boundary condition on $\phi$ reads using 
(\ref{eqpourpsi})
\begin{equation}\label{GN2}
\phi_t + (\phi u )_x + c_0 \phi_x = 0.
\end{equation}

With $c_0 = 0$, the system  (\ref{GN1}) (\ref{GN2}) is the Green-Naghdi 
system of equations \cite{Naghdi1}. The inclusion of the term $c_0\phi_x$
drastically changes its dynamics: $c_0$ cannot be eliminated  neither by a
Galilean transformation nor by a rescaling of $u$ or $\phi$. 
These extended Green-Naghdi
equations represent the nonlinear interaction between two separate forms of
motion: a wave motion associated with the elastic response of the fluid to a
perturbation,  a surface (uniform) motion generated by an external agent.

\section{Nonlinear dynamics of ripples.} 

Let us consider now asymptotic nonlinear dynamics of ripples in (\ref {GN1})
and  (\ref {GN2}). Ripples are, from a geometrical point of view, {\it nearly
local objects} and  we are looking for their {\it large time behavior}. So we
need to introduce two appropiate variables: one space variable $\zeta$
describing a local and asymptotic  pattern and a time variable $\tau$ measuring
asymptotic time dynamics. These asymptotic are worked out by means of 
the change of variables
\begin{equation}\label{varlentes}
\zeta  = \frac 1 \epsilon x\ ,\quad \tau = \epsilon t. 
\end{equation}
where the small parameter $\epsilon$ is related to the size of the wavelength:
$\ell=2\pi/k\sim\epsilon$.

Such variables cannot always be defined as shown in \cite{manna}, but exist if
the linear dispersion relation $\Omega (k)$ can be expanded as a Laurent series
with a simple pole for $k \rightarrow \infty $, that is
$$\Omega(k) = Ak + \frac{B}{k} + \frac{C}{k^3} + ....\quad k \rightarrow \infty.$$
where $A$, $B$, $C$, $...,$ are constants.
Consequently the phase and group velocities  remain bounded in the short wave limit
$k\rightarrow \infty$.  This is the case here because the dispersion relation
for (\ref{GN1}) (\ref{GN2}), obtained for $\phi (x, t) = h + a\exp
i[kx-\Omega(k) t]$ and $u(x,t)=b\exp i[kx-\Omega(k) t]$, has the asymptotic
expansion
\begin{equation}\label{reldisp}
\Omega(k) = -(\frac{3g}{c_0h})\frac {1}{ k} +\frac{3}{c_0h^2}\
(\frac{3g}{h}+\frac{3g^2}{c_0^2})\frac{1}{k^3} + {\cal O} (\frac{1}{k^5}).
\end{equation}
Such an expansion justifies the change of variables (\ref{varlentes}) and it is
essential in revealing in (\ref{GN1})  
(\ref{GN2})  the asymptotic short wave dynamics \cite{manna}. 
We are looking now for nonlinear dynamics of ripples of small
amplitude  represented by the expansions
\begin{eqnarray}
&&\phi = h + \epsilon^2 (H_0
+ \epsilon^2 H_2 + \epsilon^4 H_4 + ...)\ ,\\
&&u = \epsilon^2 (U_0 +
\epsilon^2 U_2 + \epsilon^4 U_4 + ...)\ ,
\end{eqnarray}
which from (\ref{GN1}) and (\ref{GN2}) give  at lower  order in $\epsilon$:
\begin{eqnarray}\label{GN1pert}
&&h(U_0)_\zeta + c_0(H_0)_\zeta = 0\ ,\\
\label{GN2pert}
&&g(H_0)_\zeta = \frac{h^2}{3}\{ (U_0)_{\tau \zeta} + U_0(U_0)_{\zeta
\zeta} - (U_0)_\zeta^2\}_\zeta\ . 
\end{eqnarray}
Using (\ref{GN1pert}) in (\ref{GN2pert}), going back to the 
laboratory fields ($\phi,u$) and co-ordinates ($x,t$) and integrating
once we eventually arrive at equation (\ref{mannaeq}). 

\section{Benjamin-Feir instability of the plane wave.}

For a plane wave, the nonlinears terms in (\ref{mannaeq}) give rise to
harmonics of the fundamental. Assume that a disturbance is present consisting
of modes with sideband frequences and wavenumbers close to the fundamental.  We
can have interaction between harmonics and these sideband modes.  This
interaction is likely to produce a resonant phenomenon manifesting itself by
the modulation of the plane wave solution. The exponencial growth in time of
the modulation, originating from synchronous resonance between  harmonics and
sideband modes, leads to the Benjamin-Feir instability\cite{benjaminfeir}. A
formal solutions can be given via an asymptotic expansion conducing to the
nonlinear Schodinger equation (NLS)\cite{diprima}. The particular interest of
NLS is the existence of a general and simple criterion enable to detect
stability or unstability of the monochromatic wave train. Let us seek for a
solution of (\ref{mannaeq}) under the form of a Fourier expansion in harmonics
of the fundamental  $ \exp i(kx-\Omega t)$ and where the Fourier components are
developed in a Taylor serie in powers of a small parameter $\gamma$ mesuring
the amplitude of the fundamental
\begin{equation}\label{F1}
 u = \sum_{l =-p}^{l = p}\sum_{p =1}^{\infty}  \exp il(kx -\omega t)
 \gamma^p u_{l}^p(\xi,\tau)
\end{equation} 
In (\ref{F1}), $u_{-l}^p = u_{l}^{*p}$ 
("star" denotes complex conjugation) and $\xi$ and $\tau$ are
slow variables introduced through the stretching $\xi = \gamma (x -
vt)$ and $\tau = \gamma^2 t$ and where $v$ will be determined as a
solvability condition. The expansions (\ref{F1}) includes
fast local oscillations through the dependence on the harmonics
 and slow variation (modulation) in amplitude
taken into account by the $\xi$, $\tau$ dependence of $u_{l}^{p}$.
Introducing now this expansion and the slow
variables in (\ref{mannaeq}) we may proceed to collect and solve
different order $\gamma$ and $l$. We obtain: $u_{0}^{1} = u_{2}^{2} =
u_{2}^{3} = u_{3}^{3} = 0 $, $v = \gamma /k^2 $, $u_{1}^{1} =
A(\xi,\tau) \ne 0  $ and $u_{0}^{2} = -4k^3/ \alpha $. At order $\gamma = 
3$, $l = 1$ we obtain NLS for
$A(\xi,\tau)$
\begin{equation}
iA_\tau + \frac{\alpha}{k^3} A_{\xi\xi} + \frac{4k^3}{\alpha}
A|A|^{2} = 0.
\label{NLS}
\end{equation} 
The nature of solutions of NLS depends drastically of the sign of the
product between the coeficient of $A_{\xi\xi}$ and that of $A|A|^{2}$.  
In this case the product of $ \alpha/k^3 $ by $ 4k^3/ \alpha $ 
is  positive, and according to a well known stability
criterion (see for exemple \cite{Whitham}) the plane wave solution 
(\ref{ondeplane}) is unstable.

\section{Conclusion and final coments on short waves.}

We have shown that ripples result from balance between linear
dispersion and nonlinearity as in the case of long waves.  However
the physical model under consideration represents an ideal fluid because
viscosity has been neglected, and, since dissipative phenomena take place at
small scales, viscosity must affect asymptotic dynamics of ripples.

The same approach can be used to study a flow of multiple layers. Such a flow
occurs in many practical applications (e.g multilayer coating) in which the
short wave behavior is of primordial importance.
 
Note that, while the original Green-Naghdi system  is Galilean invariant, this
invariance is lost in the Camassa-Holm equation derived by Hamiltonian methods.
In our case the perturbative theory conserves Galilean invariance and
consequently equation (\ref{mannaeq}) inherits this property from modified
Green-Naghdi.

In nonlinear dispersive systems, short wave dynamics does not occur as
naturally as long wave dynamics. It is interesting to point out that some
intermediate {\it long wave} shallow water models behave well, paradoxically,
in the short wave limit. In these intermediate models a second asymptotic limit
is always possible, in general of long wave type, leading then to the
ubiquitous KdV equation.  In some cases however, there exists a second
asymptotic limit of short wave type, leading to new nonlinear evolution
equations.  This is the case for BBMP, the integrable Camassa-Holm equation,
and one of the Boussinesq system. The short wave limit of BBMP reads
\cite{manna}: $u_{xt} = u - 3u^2$, where  $u$ has the same meaning as in
(\ref{mannaeq}).  Its solitary wave solution comes from the balance between
dispersion and nonlinearity as in KdV. Explicit solutions, blow-up and
nonlinear instabilities are studied in \cite{Silvio}.  The Camassa-Holm
equation has as short wave limit, an integrable equation which belongs to the
Harry-Dym hierarchy ($\kappa$ is a constant): $u_{xt} = \kappa u - \frac12
(u_x)^2 - uu_{xx}$.  Finally the short wave limit of the Boussinesq 
system results as :
$u_{xtt} = u_t - uu_{x}$.  The linear limit for all these cases is the equation
$u_{xt} = au$ ($a$ constant) or else 

\begin{equation}\label{swlinearlimit}
u_t -a\int_{-\infty}^x u dx = 0.
\end{equation}
This linear nonlocal equation for $u$ corresponds to the linear local
wave equation $u_t - au_x = 0$ that appears in the long wave case.
Note that the short wave limits of BBMP, Camassa-Holm and Boussinesq system  
are not evolution equations {\it stricto sensu}, rather
being integro-differential equations. 
Thus the nonlocality of nonlinear evolution equations for short waves 
appears to reflect the basic non locality of (\ref{swlinearlimit}).

\paragraph*{Aknowledgements}
The author wish to thank J. Leon and  G. Mennessier for many helpful 
and stimulating discussion.


\begin{thebibliography}{thebibliography}
\bibitem{observation}
Plane wave as a solution of a nonlinear and dispersive equation is
also known for the Nonlinear Schrodiger Equation $\psi_t + \psi_{xx}
 + |\psi|^2\psi = 0$ but $\Omega$ is $A$-dependent.
\end{thebibliography}
\end{document}